\pdfoutput=1
\documentclass[12pt,a4paper]{article}

\usepackage{ifthen}
\newboolean{pdflatex}
\setboolean{pdflatex}{true}

\newboolean{articletitles}
\setboolean{articletitles}{true}

\newboolean{uprightparticles}
\setboolean{uprightparticles}{false}

\usepackage{gensymb}
\usepackage{mathtools}
\usepackage{subcaption}
\usepackage{multicol}
\usepackage{blindtext}

\def\paperauthors{LHCb collaboration}
\def\paperasciititle{Measurement of CP violation in B0 -> psi(->ell+ell-)KS(->pi+pi-) decays}
\def\papertitle{Measurement of \CP violation in $\mbox{\decay{\Bz}{\psi(\to\ellp\ellm)\KS(\to\pip\pim)}}$ decays}
\def\paperkeywords{{High Energy Physics}, {LHCb}}
\def\papercopyright{\the\year\ CERN for the benefit of the LHCb collaboration}
\def\paperlicence{CC BY 4.0 licence}
\def\paperlicenceurl{https://creativecommons.org/licenses/by/4.0/}

\usepackage[top=1in, bottom=1.25in, left=1in, right=1in]{geometry}

\usepackage{microtype}
\usepackage{csquotes}
\usepackage{csquotes}
\usepackage{xspace}
\usepackage{caption}
\usepackage{subcaption}
\usepackage{lineno}

\usepackage{placeins}

\usepackage{graphicx}
\usepackage{color}
\usepackage{colortbl}
\graphicspath{{./figs/}}

\usepackage{amsmath}
\usepackage{amssymb}
\usepackage{booktabs}
\usepackage{amsfonts}
\usepackage{upgreek}

\newcommand*\patchAmsMathEnvironmentForLineno[1]{%
\expandafter\let\csname old#1\expandafter\endcsname\csname #1\endcsname
\expandafter\let\csname oldend#1\expandafter\endcsname\csname
end#1\endcsname
 \renewenvironment{#1}%
   {\linenomath\csname old#1\endcsname}%
   {\csname oldend#1\endcsname\endlinenomath}%
}
\newcommand*\patchBothAmsMathEnvironmentsForLineno[1]{%
  \patchAmsMathEnvironmentForLineno{#1}%
  \patchAmsMathEnvironmentForLineno{#1*}%
}
\AtBeginDocument{%
\patchBothAmsMathEnvironmentsForLineno{equation}%
\patchBothAmsMathEnvironmentsForLineno{align}%
\patchBothAmsMathEnvironmentsForLineno{flalign}%
\patchBothAmsMathEnvironmentsForLineno{alignat}%
\patchBothAmsMathEnvironmentsForLineno{gather}%
\patchBothAmsMathEnvironmentsForLineno{multline}%
\patchBothAmsMathEnvironmentsForLineno{eqnarray}%
}

\usepackage{hyperxmp}

\usepackage[pdftex,
            pdfauthor={\paperauthors},
            pdftitle={\paperasciititle},
            pdfkeywords={\paperkeywords},
            pdfcopyright={Copyright (C) \papercopyright},
            pdflicenseurl={\paperlicenceurl}]{hyperref}
\usepackage[colorinlistoftodos,textsize=scriptsize]{todonotes}

\usepackage[bottom,flushmargin,hang,multiple]{footmisc}

\usepackage[all]{hypcap}

\usepackage{xspace} 
\usepackage{upgreek}

\def\lhcb   {\mbox{LHCb}\xspace}

\def\belle  {\mbox{Belle}\xspace}
\def\belletwo {\mbox{Belle~II}\xspace}

\def\velo   {VELO\xspace}

\def\MagUp {\mbox{\em Mag\kern -0.05em Up}\xspace}

\ifthenelse{\boolean{uprightparticles}}%
{

 \def\Pmu         {\ensuremath{\upmu}\xspace}

 \def\Ppi         {\ensuremath{\uppi}\xspace}

 \def\Ppsi        {\ensuremath{\uppsi}\xspace}

 \def\PDelta      {\ensuremath{\Delta}\xspace}                 
 \def\PXi         {\ensuremath{\Xi}\xspace}                 
 \def\PLambda     {\ensuremath{\Lambda}\xspace}                 
 \def\PSigma      {\ensuremath{\Sigma}\xspace}                 
 \def\POmega      {\ensuremath{\Omega}\xspace}                 
 \def\PUpsilon    {\ensuremath{\Upsilon}\xspace}
 \let\oldPi\Pi
 \def\PPi         {\ensuremath{\oldPi}\xspace}

 \def\PB      {\ensuremath{\mathrm{B}}\xspace}                 
                  
 \def\PD      {\ensuremath{\mathrm{D}}\xspace}

 \def\PJ      {\ensuremath{\mathrm{J}}\xspace}                 
 \def\PK      {\ensuremath{\mathrm{K}}\xspace}

 \def\PP      {\ensuremath{\mathrm{P}}\xspace}

 \def\Pb      {\ensuremath{\mathrm{b}}\xspace}                 
 \def\Pc      {\ensuremath{\mathrm{c}}\xspace}                 
 \def\Pd      {\ensuremath{\mathrm{d}}\xspace}                 
 \def\Pe      {\ensuremath{\mathrm{e}}\xspace}

 \def\Pi      {\ensuremath{\mathrm{i}}\xspace}

 \def\Pp      {\ensuremath{\mathrm{p}}\xspace}

 \def\Ps      {\ensuremath{\mathrm{s}}\xspace}

 \def\thebaroffset{0.0em}
}
{

 \def\Pmu         {\ensuremath{\mu}\xspace}

 \def\Ppi         {\ensuremath{\pi}\xspace}

 \def\Ppsi        {\ensuremath{\psi}\xspace}                 
                  
 \mathchardef\PDelta="7101
 \mathchardef\PXi="7104
 \mathchardef\PLambda="7103
 \mathchardef\PSigma="7106
 \mathchardef\POmega="710A
 \mathchardef\PUpsilon="7107
 \mathchardef\PPi="7105
                  
 \def\PB      {\ensuremath{B}\xspace}                 
                  
 \def\PD      {\ensuremath{D}\xspace}

 \def\PJ      {\ensuremath{J}\xspace}                 
 \def\PK      {\ensuremath{K}\xspace}

 \def\PP      {\ensuremath{P}\xspace}

 \def\Pb      {\ensuremath{b}\xspace}                 
 \def\Pc      {\ensuremath{c}\xspace}                 
 \def\Pd      {\ensuremath{d}\xspace}                 
 \def\Pe      {\ensuremath{e}\xspace}

 \def\Pi      {\ensuremath{i}\xspace}

 \def\Pp      {\ensuremath{p}\xspace}

 \def\Ps      {\ensuremath{s}\xspace}

 \def\thebaroffset{0.18em}
}
\newcommand{\offsetoverline}[2][\thebaroffset]{\kern #1\overline{\kern -#1 #2}}%

\makeatletter
\ifcase \@ptsize \relax
  \newcommand{\miniscule}{\@setfontsize\miniscule{4}{5}}
\or
  \newcommand{\miniscule}{\@setfontsize\miniscule{5}{6}}
\or
  \newcommand{\miniscule}{\@setfontsize\miniscule{5}{6}}
\fi
\makeatother

\DeclareRobustCommand{\optbar}[1]{\shortstack{{\miniscule (\rule[.5ex]{1.25em}{.18mm})}
  \\ [-.7ex] $#1$}}

\def\electron   {{\ensuremath{\Pe}}\xspace}
\def\en         {{\ensuremath{\Pe^-}}\xspace}   
\def\ep         {{\ensuremath{\Pe^+}}\xspace}

\def\mup        {{\ensuremath{\Pmu^+}}\xspace}
\def\mun        {{\ensuremath{\Pmu^-}}\xspace} 

\def\ellm       {{\ensuremath{\ell^-}}\xspace}
\def\ellp       {{\ensuremath{\ell^+}}\xspace}

\def\dquark    {{\ensuremath{\Pd}}\xspace}

\def\squark    {{\ensuremath{\Ps}}\xspace}

\def\cquark    {{\ensuremath{\Pc}}\xspace}
\def\cquarkbar {{\ensuremath{\overline \cquark}}\xspace}

\def\bquark    {{\ensuremath{\Pb}}\xspace}

\def\pion   {{\ensuremath{\Ppi}}\xspace}

\def\pip    {{\ensuremath{\pion^+}}\xspace}
\def\pim    {{\ensuremath{\pion^-}}\xspace}
\def\pipm   {{\ensuremath{\pion^\pm}}\xspace}

\def\kaon    {{\ensuremath{\PK}}\xspace}
\def\Kbar    {{\ensuremath{\offsetoverline{\PK}}}\xspace}

\def\KorKbar {\kern \thebaroffset\optbar{\kern -\thebaroffset \PK}{}\xspace}
\def\Kz      {{\ensuremath{\kaon^0}}\xspace}
\def\Kzb     {{\ensuremath{\Kbar{}^0}}\xspace}
\def\Kp      {{\ensuremath{\kaon^+}}\xspace}

\def\KS      {{\ensuremath{\kaon^0_{\mathrm{S}}}}\xspace}

\def\Kstarz  {{\ensuremath{\kaon^{*0}}}\xspace}

\def\Kstar   {{\ensuremath{\kaon^*}}\xspace}

\def\D       {{\ensuremath{\PD}}\xspace}

\def\DorDbar {\kern \thebaroffset\optbar{\kern -\thebaroffset \PD}\xspace}

\def\Dp      {{\ensuremath{\D^+}}\xspace}
\def\Dm      {{\ensuremath{\D^-}}\xspace}

\def\DpDm    {\ensuremath{\Dp {\kern -0.16em \Dm}}\xspace}

\def\B       {{\ensuremath{\PB}}\xspace}
\def\Bbar    {{\ensuremath{\offsetoverline{\PB}}}\xspace}

\def\BorBbar {\kern \thebaroffset\optbar{\kern -\thebaroffset \PB}\xspace}
\def\Bz      {{\ensuremath{\B^0}}\xspace}
\def\Bzb     {{\ensuremath{\Bbar{}^0}}\xspace}
\def\Bd      {{\ensuremath{\B^0}}\xspace}

\def\BdorBdbar {\kern \thebaroffset\optbar{\kern -\thebaroffset \Bd}\xspace}
\def\Bu      {{\ensuremath{\B^+}}\xspace}

\def\Bp      {{\ensuremath{\Bu}}\xspace}

\def\Bs      {{\ensuremath{\B^0_\squark}}\xspace}

\def\BsorBsbar {\kern \thebaroffset\optbar{\kern -\thebaroffset \Bs}\xspace}

\def\jpsi     {{\ensuremath{{\PJ\mskip -3mu/\mskip -2mu\Ppsi}}}\xspace}
\def\psitwos  {{\ensuremath{\Ppsi{(2S)}}}\xspace}

\def\Y#1S{\ensuremath{\PUpsilon{(#1S)}}\xspace}

\def\proton      {{\ensuremath{\Pp}}\xspace}

\def\Lz          {{\ensuremath{\PLambda}}\xspace}

\def\LorLbar     {\kern \thebaroffset\optbar{\kern -\thebaroffset \PLambda}\xspace}

\def\Lb           {{\ensuremath{\Lz^0_\bquark}}\xspace}

\newcommand{\decay}[2]{\ensuremath{#1\!\to #2}\xspace} 

\def\to                 {\ensuremath{\rightarrow}\xspace}

\def\CP                {{\ensuremath{C\!P}}\xspace}

\newcommand{\dmd}{{\ensuremath{\Delta m_{\dquark}}}\xspace}
\newcommand{\DG}{{\ensuremath{\Delta\Gamma}}\xspace}

\newcommand{\DGd}{{\ensuremath{\Delta\Gamma_{\dquark}}}\xspace}

\newcommand{\ACP}{{\ensuremath{{\mathcal{A}}^{\CP}}}\xspace}

\newcommand{\phid}{{\ensuremath{\phi_{\dquark}}}\xspace}

\def\AT#1     {\ensuremath{A_{\mathrm{T}}^{#1}}\xspace}           

\def\C#1      {\ensuremath{\mathcal{C}_{#1}}\xspace}                       
\def\Cp#1     {\ensuremath{\mathcal{C}_{#1}^{'}}\xspace}                    
\def\Ceff#1   {\ensuremath{\mathcal{C}_{#1}^{\mathrm{(eff)}}}\xspace}        
\def\Cpeff#1  {\ensuremath{\mathcal{C}_{#1}^{'\mathrm{(eff)}}}\xspace}       
\def\Ope#1    {\ensuremath{\mathcal{O}_{#1}}\xspace}                       
\def\Opep#1   {\ensuremath{\mathcal{O}_{#1}^{'}}\xspace}                    

\newcommand{\aunit}[1]{\ensuremath{\text{\,#1}}}       

\newcommand{\tev}{\aunit{Te\kern -0.1em V}\xspace}
\newcommand{\gev}{\aunit{Ge\kern -0.1em V}\xspace}
\newcommand{\mev}{\aunit{Me\kern -0.1em V}\xspace}
\newcommand{\kev}{\aunit{ke\kern -0.1em V}\xspace}
\newcommand{\ev}{\aunit{e\kern -0.1em V}\xspace}
 
\newcommand{\mevc}{\ensuremath{\aunit{Me\kern -0.1em V\!/}c}\xspace}
\newcommand{\gevc}{\ensuremath{\aunit{Ge\kern -0.1em V\!/}c}\xspace}
\newcommand{\mevcc}{\ensuremath{\aunit{Me\kern -0.1em V\!/}c^2}\xspace}
\newcommand{\gevcc}{\ensuremath{\aunit{Ge\kern -0.1em V\!/}c^2}\xspace}

\def\fb   {\ensuremath{\aunit{fb}}\xspace}
\def\invfb   {\ensuremath{\fb^{-1}}\xspace}

\def\ps   {\ensuremath{\aunit{ps}}\xspace}
\def\fs   {\aunit{fs}}

\def\gsim{{~\raise.15em\hbox{$>$}\kern-.85em
          \lower.35em\hbox{$\sim$}~}\xspace}
\def\lsim{{~\raise.15em\hbox{$<$}\kern-.85em
          \lower.35em\hbox{$\sim$}~}\xspace}

\def\pt         {\ensuremath{p_{\mathrm{T}}}\xspace}

\def\evtgen     {\mbox{\textsc{EvtGen}}\xspace}

\def\geant      {\mbox{\textsc{Geant4}}\xspace}

\def\photos     {\mbox{\textsc{Photos}}\xspace}

\def\pythia     {\mbox{\textsc{Pythia}}\xspace}

\def\tell1  {TELL1\xspace}
\def\ukl1   {UKL1\xspace}

\newcommand{\lhcborcid}[1]{\href{https://orcid.org/#1}{\hspace*{0.1em}\raisebox{-0.45ex}{\includegraphics[width=1em]{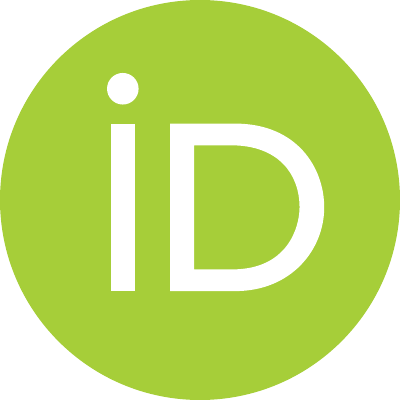}}}}

\usepackage{cite}
\usepackage{mciteplus}

\newcommand{\runone}{\text{Run 1 }}
\newcommand{\runtwo}{\text{Run 2 }}
\newcommand{\runonetwo}{\text{Run 1\&2 }}
\newcommand{\BPsiKS}{\decay{\Bz}{\psi\KS}}
\newcommand{\sintwobeta}{\ensuremath{\sin(2\beta)}}

\usepackage[
  locale=US,
  separate-uncertainty=true,
  per-mode=symbol-or-fraction,
]{siunitx}
\usepackage{cleveref}

\begin{document}

\renewcommand{\thefootnote}{\fnsymbol{footnote}}
\setcounter{footnote}{1}


\begin{titlepage}
\pagenumbering{roman}

\vspace*{-1.5cm}
\centerline{\large EUROPEAN ORGANIZATION FOR NUCLEAR RESEARCH (CERN)}
\vspace*{1.5cm}
\noindent
\begin{tabular*}{\linewidth}{lc@{\extracolsep{\fill}}r@{\extracolsep{0pt}}}
\ifthenelse{\boolean{pdflatex}}
{\vspace*{-1.5cm}\mbox{\!\!\!\includegraphics[width=.14\textwidth]{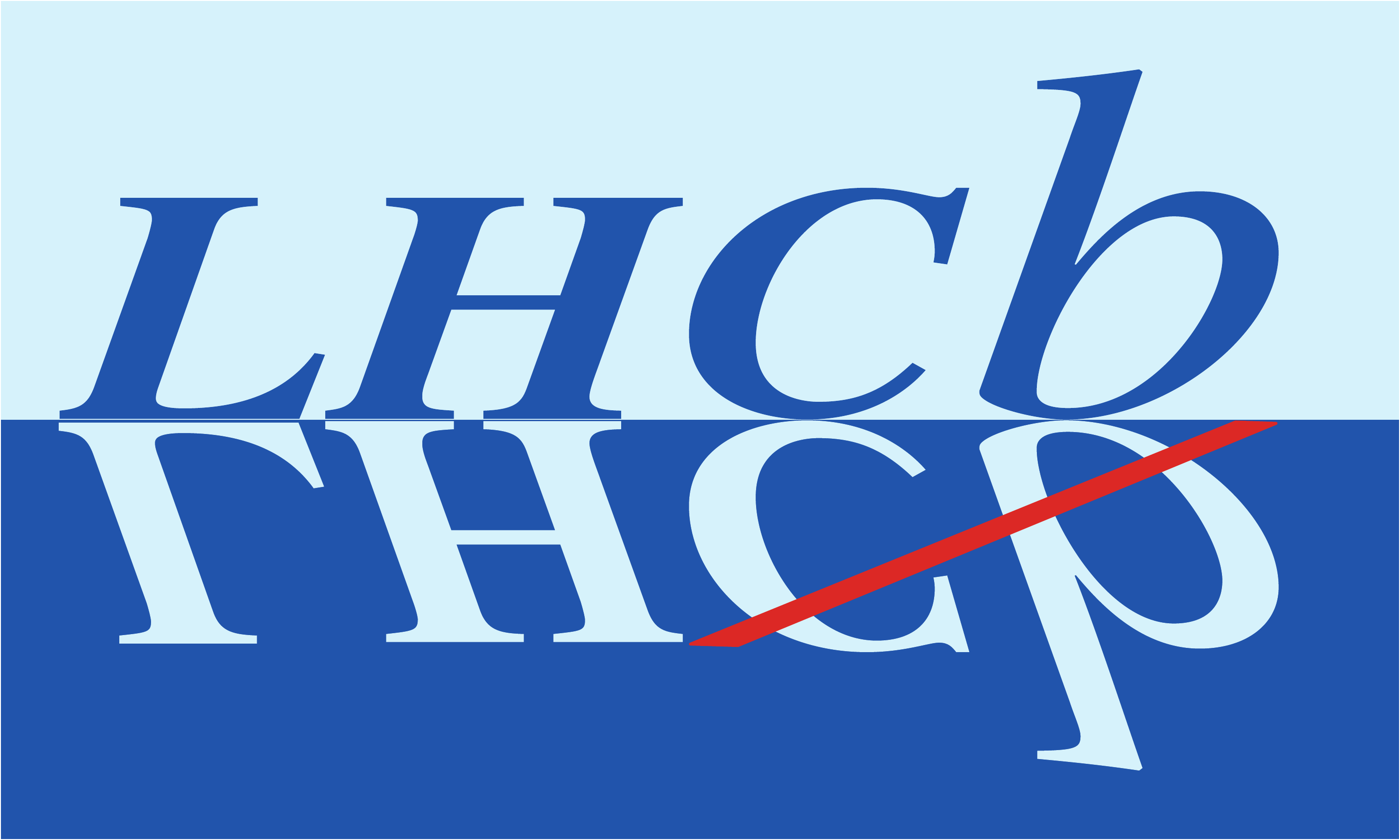}} & &}%
{\vspace*{-1.2cm}\mbox{\!\!\!\includegraphics[width=.12\textwidth]{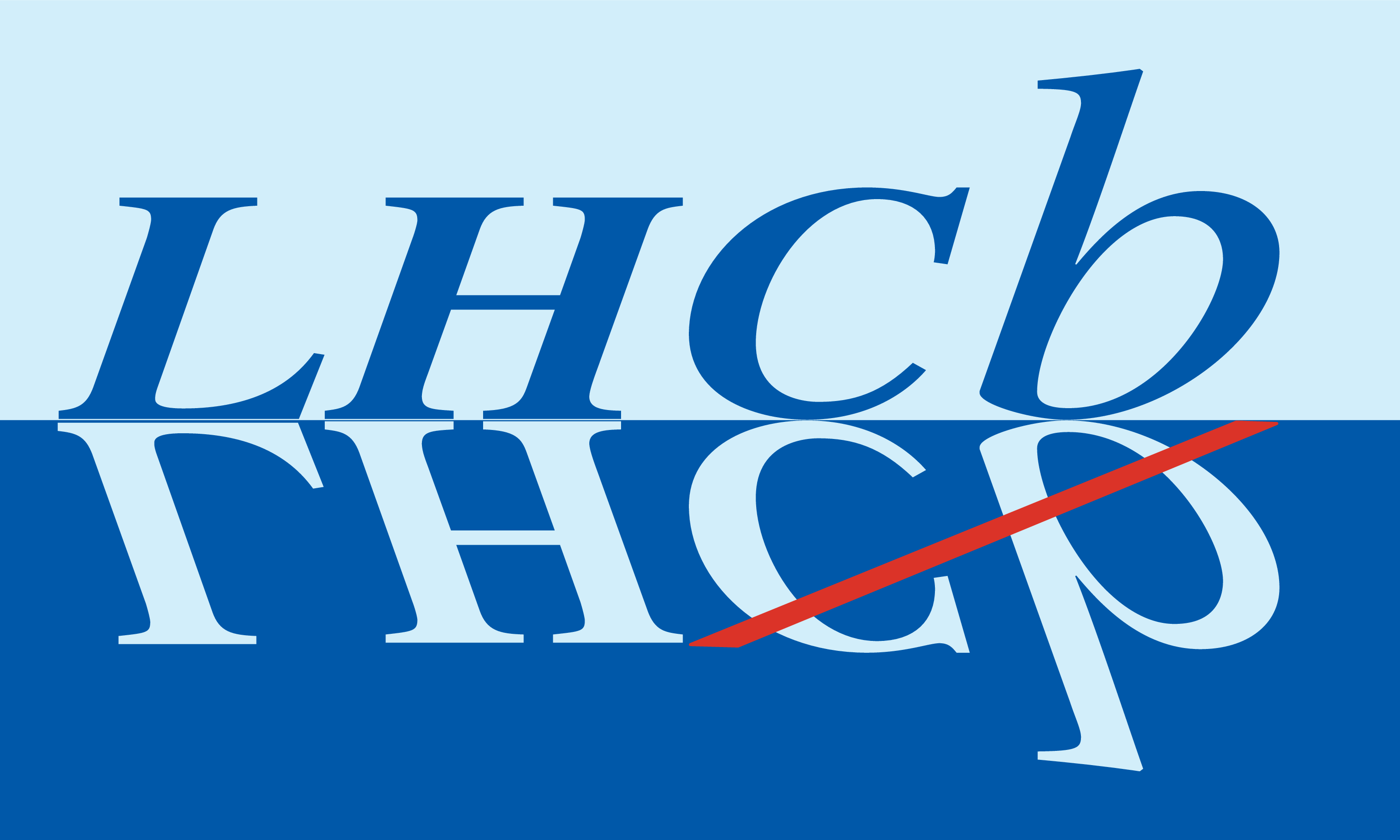}} & &}%
\\
 & & CERN-EP-2023-177 \\  
 & & LHCb-PAPER-2023-013 \\  
 & & 9 January 2024 \\
 & & \\
\end{tabular*}

\vspace*{3.0cm}

{\normalfont\bfseries\boldmath\huge
\begin{center}
  \papertitle 
\end{center}
}

\vspace*{1.5cm}

\begin{center}
\paperauthors\footnote{Authors are listed at the end of this Letter.}
\end{center}

\vspace{\fill}

\begin{abstract}
  \noindent
  A measurement of time-dependent $\CP$~violation in the decays of \Bz and \Bzb mesons to the final states
  ${\jpsi(\to\mup\mun)\KS}$, $\psitwos(\to\mup\mun)\KS$ and
  $\jpsi(\to\ep\en)\KS$ with \decay{\KS}{\pip\pim} is presented. The data
  correspond to an integrated luminosity of
  $6\invfb$ collected at a center-of-mass energy of $\sqrt{s}=13\tev$ with the
  \lhcb detector. The $\CP$-violation parameters are measured to be
  \begin{align*}
  S_{\psi\KS} &= 0.717 \pm 0.013\,(\text{stat}) \pm 0.008\,(\text{syst}), \\
  C_{\psi\KS} &= 0.008 \pm 0.012\,(\text{stat}) \pm 0.003\,(\text{syst}).
  \end{align*}
  This measurement of $S_{\psi\KS}$ represents the most precise single measurement of the CKM angle $\beta$ to date and is more precise than the current world average.
  In addition, measurements of the \CP-violation parameters of the individual channels are reported and a combination with the \lhcb \runone measurements is performed.
\end{abstract}

\vspace*{1.5cm}

\begin{center}
    Published in Phys.~Rev.~Lett. 132 (2024) 021801
\end{center}

\vspace{\fill}

{\footnotesize 
\centerline{\copyright~\papercopyright. \href{\paperlicenceurl}{\paperlicence}.}}
\vspace*{2mm}

\end{titlepage}


\newpage
\setcounter{page}{2}
\mbox{~}


\renewcommand{\thefootnote}{\arabic{footnote}}
\setcounter{footnote}{0}

\cleardoublepage


\pagestyle{plain}
\setcounter{page}{1}
\pagenumbering{arabic}

Charge-parity ($\CP$) symmetry violation in transitions of neutral mesons to \CP-invariant final states can occur in the interference of the decay amplitudes with and without neutral meson mixing.
The \BPsiKS{} decays involving charmonium states, $\psi=[\jpsi,\psitwos]$, are dominated by the \decay{\bquark}{\cquark\cquarkbar\squark} tree-level transition and have a clean experimental signature.
Measurements of \CP violation in neutral meson decays to charmonium final states have thus resulted in a high degree of precision for the angle $\beta$ of the Cabibbo Kobayashi Maskawa (CKM) matrix: ${\sintwobeta=0.699\pm 0.017}$~\cite{HFLAV21}.
The first observation of \CP violation in the \PB-meson system was reported in the \decay{\Bz}{\jpsi\KS} channel  by the \emph{BABAR}~\cite{PhysRevLett.87.091801} and \belle~\cite{PhysRevLett.87.091802} collaborations.
The measurement of the \CP-violation parameter \sintwobeta{} has been updated several times by these experiments~\cite{BaBar:2009byl,Belle:2012paq} and more recently by the \lhcb~\cite{LHCb-PAPER-2015-004,LHCb-PAPER-2017-029} and \belletwo\cite{Belle-II:2023nmj} collaborations. 

A measurement of $\CP$ violation in the interference of \Bz decays with and without mixing entails the measurement of the periodic change of decay-rate differences of initial $\Bz$ and $\Bzb$ mesons to a common, \CP-invariant final state, $f$, with time.
The underlying decay-time-dependent $\CP$ asymmetry can be expressed as
\begin{equation}
    \ACP(t) = \frac{\Gamma(\decay{\Bzb(t)}{f}) - \Gamma(\decay{\Bz(t)}{f})}{\Gamma(\decay{\Bzb(t)}{f}) + \Gamma(\decay{\Bz(t)}{f})} = \frac{S\sin(\dmd t)-C\cos(\dmd t)}{\cosh\!\left(\frac{1}{2}\DGd t\right)+\mathcal{A}_\DG\!\sinh\!\left(\frac{1}{2}\DGd t\right)},\label{eqn:asymmetry}
\end{equation}
where $S$, $C$, and $\mathcal{A}_\DG$ are the $\CP$-violation parameters and \dmd is the \Bz-\Bzb mixing frequency~\cite{CPreviewChap}.
The parameter $S$ can be related to the CKM angle $\beta$ as ${S=\sin(2\beta+\Delta\phid+\Delta\phid^\text{NP})}$, where $\Delta\phid^\text{NP}$ is a possible contribution from physics beyond the standard model.
Contributions from penguin topologies to the decay amplitude that cause an additional phase shift $\Delta\phid$ are CKM suppressed, hence deviations of $S$ from $\sin(2\beta)$ are expected to be small in the standard model~\cite{DeBruyn:2014oga,Barel:2020jvf,Ciuchini:2005mg,Faller:2008zc,Jung:2012mp,Frings:2015eva}. 
In this measurement it is assumed that the width difference between \PB mass eigenstates, \DGd, is compatible with zero~\cite{HFLAV21}, such that the denominator on the right-hand side of Eq.~\eqref{eqn:asymmetry} is equal to one.

The \lhcb detector~\cite{LHCb-DP-2008-001,LHCb-DP-2014-002} is a single-arm spectrometer covering the forward pseudorapidity region, designed for the study of particles containing \bquark\ or \cquark\ quarks.
The detector elements that are particularly relevant to this measurement are a silicon-strip vertex locator (\velo) surrounding the proton-proton ($\proton\proton$) interaction region that allows \cquark\ and \bquark\ hadrons to be identified from their characteristically long flight distance; a tracking system that consists of a dipole magnet and sets of tracking stations before and after the magnet, and provides a measurement of the momentum, $p$, of charged particles; and two ring-imaging Cherenkov detectors (RICH) that are able to discriminate between different species of charged hadrons.
In addition, a muon system allows the identification of muons, and a calorimeter system provides electron identification.
The online event selection is performed by a trigger~\cite{LHCb-DP-2012-004}, which consists of a hardware stage, based on information from the calorimeter and muon systems, followed by a software stage, which applies a full event reconstruction.

The measurements reported in this Letter use the $\proton\proton$ collision data collected by the \lhcb experiment during \runtwo (2015 to 2018) at a center-of-mass energy of 13\tev.
The three decay modes  ${\decay{\Bz}{\jpsi(\to\mup\mun)\KS}}$, ${\decay{\Bz}{\psitwos(\to\mup\mun)\KS}}$, and ${\decay{\Bz}{\jpsi(\to\ep\en)\KS}}$, with \decay{\KS}{\pip\pim}, are studied, and a combined, time-dependent maximum-likelihood fit is performed to measure the \CP-violation parameters.
To select the signal modes, trigger decisions associated with particles reconstructed off-line are applied, and requirements are made on whether the decision was caused by the reconstructed signal candidate, by other reconstructed particles in the event, or both.
For the dimuon channels, only candidates responsible for the muon trigger decision are considered.
For the reconstruction of \decay{\psi}{\mup\mun} candidates, muons with momentum transverse to the beam direction, $\pt$, larger than $500\mevc$ are required to form a good-quality vertex.
The dimuon invariant mass must be within $\pm 100\mevcc$ of the known \jpsi or \psitwos masses~\cite{PDG2022}.
For the selection of \decay{\jpsi}{\ep\en} candidates, all events selected by the trigger are considered.
The two electron candidates are required to have $\pt>500\mevc$ and to form a good-quality vertex.
The dielectron invariant mass is required to be in the range $[2300,3300]\mevcc$.
The final-state \KS meson is reconstructed from two tracks forming a good-quality vertex.

Because of its long flight distance, about two thirds of all $\decay{\KS}{\pip\pim}$ decays occur downstream of the \velo but before the next tracking subdetector, the tracker turicensis (TT), which is located upstream of the \lhcb magnet.
Tracks that are reconstructed from hits in the \velo as well as by subsequent \lhcb tracking detectors are called long (\emph{L}).
Tracks that are only measured by the TT and the tracking stations are called downstream (\emph{D}).
Consequently a \KS candidate can be reconstructed as a \enquote{\emph{LL}}, \enquote{\emph{DD}}, or \enquote{\emph{LD}} track combination.
In addition, upstream (\emph{U}) tracks, which are only measured by the \velo and TT and then bent out of the \lhcb acceptance due to their low momentum, are included and matched to a \emph{L} track to form a \enquote{\emph{UL}} \KS candidate.
\emph{L} and \emph{D} tracks are required to have momenta above $2\gevc$ whereas for \emph{U} tracks the requirement is lower, at $1\gevc$.
Moreover, the two tracks are required to have a reconstructed invariant mass within $\pm 80\mevcc$ of the known \KS mass~\cite{PDG2022}.
The \KS reconstruction categories \emph{LD} and \emph{UL} are included for the first time in a measurement of time-dependent \CP violation at \lhcb and add \SI{13}{\percent} to the signal yield of the $\decay{\Bz}{\jpsi(\to\mup\mun)\KS}$ mode.
As the precision of the measured dimuon vertex  determines the decay-time precision and does not depend on the kaon reconstruction, the new reconstruction categories were found to be as well suited for the measurement as the \emph{LL} and \emph{DD} categories.
The signal \Bz candidates are formed from reconstructed \KS and $\psi$ candidates forming a good-quality  vertex, which is displaced from the primary $\proton\proton$ interaction vertex (PV).
Because of the inclusive trigger approach used for the dielectron channel, an additional off-line selection is applied to reduce the background level.
Reduced \KS mass windows of $25\mevcc$ and $33\mevcc$ are chosen for \emph{LL} and \emph{DD} \KS candidates, and a more stringent requirement on the alignment between the \PB momentum vector and the direction defined by the PV and \PB decay vertex is applied.

Background due to ${\decay{\Lb}{\psi\Lz(\to\proton\pim)}}$ decays where the proton is misidentified as a pion is suppressed by removing candidates with the recalculated proton-pion invariant mass in the range $[1110, 1130]\mevcc$ and a poor pion identification likelihood, as measured by the RICH detectors.
Background contributions from decays involving short-lived intermediate hadronic resonances like $\decay{\Bz}{\psi\Kstarz}$ with a misidentified charged kaon are suppressed by requiring that the reconstructed \KS{} decay time exceeds $\num{0.5}\ps$.
In the following, \Kstarz denotes the resonance $\Kstar(892)^0$.
The $\decay{\Bp}{\psi\Kp}$ decay can also mimic the signal if an extra low-momentum \emph{U}-track pion is combined with the misidentified charged kaon.
This source of background contribution is reduced by requiring that the pion candidates have low kaon-identification likelihoods.

The background due to random track combinations, named combinatorial background, is reduced with a boosted decision tree classifier (BDT)~\cite{XGBoost}.
This model is trained on simulated \BPsiKS{} decays as the signal proxy and \PB-meson candidates with $m(\psi\KS)>5400\mevcc$ as the background proxy.
In the simulation, the $\proton\proton$ collisions are generated with the event-generator \pythia~\cite{Sjostrand:2007gs,*Sjostrand:2006za} with a specific \lhcb configuration~\cite{LHCb-PROC-2010-056}.
The material interaction in the detector is simulated with the \geant toolkit~\cite{Allison:2006ve, *Agostinelli:2002hh}.
Decays of unstable particles are simulated with \evtgen \cite{Lange:2001uf} with a phase-space model including \CP violation, in which final-state radiation is generated using \photos~\cite{davidson2015photos}.
The training features show good agreement between background-subtracted data and simulation.
They include the impact parameter of the reconstructed \Bz, $\psi$, and \KS candidates, defined as the shortest distance between each particle's reconstructed trajectory and the PV, the \pt of the \KS and \pipm mesons and the \KS pseudorapidity.
A kinematic fit to the signal decay tree where the $\psi$ and \KS masses are constrained to their known values is performed, and the resulting fit $\chi^2$ is added to the set of training features.
To minimize overtraining, a $k$-folding technique~\cite{Blum1999BeatingTH} is used with $k=5$ and the training is stopped as soon as the classifiers' separation power no longer significantly improves on the test sample.
The predictions of the BDTs trained on the five folds are averaged.
The BDT classifiers are trained separately for each \KS reconstruction class.
The selection is finalized by applying a requirement on the BDT output that maximizes the signal yield sensitivity.
Following the selection procedure, \SI{0.7}{\percent} of events contain multiple signal candidates; in these cases the candidate with the highest \pt is selected.

The flavor of the \PB meson at production, needed to evaluate Eq.~\eqref{eqn:asymmetry}, is determined by different methods that either exploit additional particles produced in the fragmentation of the \bquark quark associated with the signal-\PB meson (same side tagging, SS) or from the decay of the \bquark hadron produced in association with the signal (opposite side tagging, OS).
The particles considered in these processes include protons and pions from the signal \bquark-quark fragmentation process, and kaons, electrons, muons and reconstructed charged charm-hadron decays as well as a weighted mean of the charges of all reconstructed decay vertices of the associated partner \bquark-hadron decay.
Collectively, these methods are referred to as flavor tagging (FT)~\cite{LHCb-PAPER-2011-027,Fazzini:2018dyq}.
From the measured particle charges, the initial \PB state can be inferred in the form of a tagging decision ${d=(+1,-1)\equiv(\Bz,\Bzb)}$.
In addition, a probability estimate that the assigned decision is wrong (predicted mistag or $\eta$) is estimated for each candidate by means of a multivariate classification method.

The predicted mistag is calibrated by means of a dedicated software package~\cite{ftcalib}, using flavor-specific channels that are kinematically similar to the signal channels, to guarantee the calibrated mistag function, $\omega(\eta)$, closely matches the mistag probability in the calibration channels.
The difference in the tagging response for \Bzb and \Bz mesons is taken into account with separate calibrated tagging responses $\omega^-(\eta)$ and $\omega^+(\eta)$, respectively.
Two calibration channels are considered in this analysis: the \decay{\Bp}{\jpsi\Kp} and \decay{\Bz}{\jpsi\Kstarz(\to\!\Kp\pim)} decays where the \jpsi is either reconstructed from two electrons or two muons, depending on the targeted signal mode.
Selection criteria similar to the signal requirements are applied and weights to subtract background contributions are determined from a fit to the \PB candidates invariant mass distribution with the sPlot method~\cite{Pivk:2004ty,Xie:2009rka}.
Before calibrating the tagging output, the samples are weighted such that the relevant candidate kinematic distributions and properties match those of the corresponding \PB candidates in the signal decay modes.
These distributions and properties are the \PB candidate pseudorapidity, \pt and azimuth angle, and the number of reconstructed PVs and tracks in each event.
After the calibration of individual tagging algorithms, the OS and SS combinations are calibrated using both control channels.

The final FT calibration parameters are determined from a fit to the decay-time distribution of the \decay{\Bz}{\jpsi\Kstarz} signal candidates where constraints on the OS parameters from the \decay{\Bp}{\jpsi\Kp}  calibration are applied.
In this fit the asymmetries of the \Bz-\Bzb production, $\alpha\equiv[N(\Bzb)-N(\Bz))/(N(\Bzb)+N(\Bz)]$ and reconstruction, as well as of tagging efficiencies, defined here as $\Delta\epsilon_{\rm tag}\equiv (\epsilon^\Bzb_\text{tag}-\epsilon^\Bz_\text{tag})/(\epsilon^\Bzb_\text{tag}+\epsilon^\Bz_\text{tag})$, are determined and production and tagging asymmetries are propagated to the signal fit.
The tagging power measures the effective loss in signal yield compared to a perfectly tagged sample for a measurement of the time-dependent \CP asymmetry.
It is calculated as $\epsilon_\text{tag}\mathcal{D}^2$, where the tagging efficiency $\epsilon_\text{tag}$ is the fraction of tagged events in relation to the total sample size and $\mathcal{D}$ is the FT dilution factor $[1-2\omega(\eta)]$.
A summary of the values of $\epsilon_\text{tag}$ and $\mathcal{D}^2$ found for each considered decay is given in Table\ \ref{tab:ftperf}, and in the following, only events with available tagging decisions are considered.
The tagging power depends on the selection criteria used to isolate each signal.
For the dielectron mode, the stricter requirements imposed by the trigger and subsequent selection cause the tagging power to be higher than for the dimuon modes.
\begin{table}
    \centering
    \caption{Flavor tagging efficiency and $\mathcal{D}^2$ factor for each decay channel.}
    \label{tab:ftperf}
    \begin{tabular}{lS[table-format=2.2]@{$\,\pm\,$}S[table-format=1.2]S[table-format=1.3]@{$\,\pm\,$}S[table-format=1.3]}
         \toprule
         Channel & \multicolumn{2}{c}{$\epsilon_\text{tag}\,[\%]$} & \multicolumn{2}{c}{$\mathcal{D}^2\,[\%]$} \\
         \midrule
         {\small\decay{\Bz}{\jpsi(\to\mup\mun)\KS}}     & 85.34 & 0.05 & 4.661 & 0.013 \\
         {\small\decay{\Bz}{\jpsi(\to\ep\en)\KS}}       & 92.20 & 0.08 & 6.462 & 0.032 \\
         {\small\decay{\Bz}{\psitwos(\to\mup\mun)\KS}}  & 84.81 & 0.15 & 4.59  & 0.04  \\
         \bottomrule
    \end{tabular}
\end{table}

Unbinned, extended maximum-likelihood fits to the invariant-mass distributions of the signal candidates are performed for each final state to determine the signal and background contributions.
From the result of these fits, sets of weights are determined using the sPlot method and are used to obtain the signal decay-time distributions from the data.
The \Bz signal is described by a two-sided Hypatia probability density function (PDF)~\cite{Santos:2013gra} in each channel.
The Hypatia parameters defining the tails are determined from the respective simulation sample.
The width and mean are allowed to float in the fit to the data.
The same model and all its shape parameters are used to describe background \Bs decays into the same final states, but the mean, relative to that of the \Bz component, is offset by the known mass difference~\cite{PDG2022}.
The combinatorial background is described by an exponential distribution and the partially reconstructed low-mass background is described by a Gaussian distribution.

\begin{figure}[t]
    \centering
  \includegraphics[width=8.6cm]{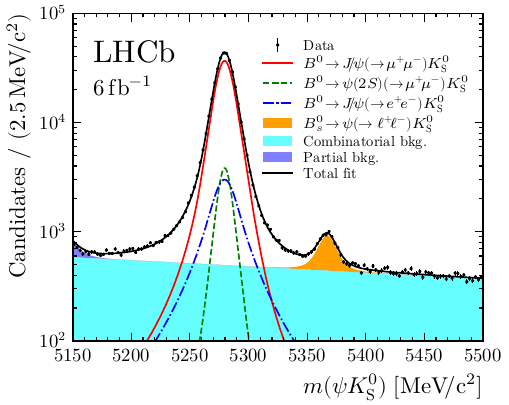}
  \caption{
    Invariant-mass distribution of the selected candidates with an identified flavor at production of the three signal channels.}
  \label{fig:massfits}
\end{figure}

\Cref{fig:massfits} shows the invariant mass distribution of the selected candidates of all the decay modes, with the full fit functions and their partial contributions.
In total, \num{306090+-570}, \num{23560+-160}, and \num{42700+-220} signal decays with an identified flavor at production are found in the modes ${\decay{\Bz}{\jpsi(\to\mup\mun)\KS}}$, ${\decay{\Bz}{\psitwos(\to\mup\mun)\KS}}$ and ${\decay{\Bz}{\jpsi(\to\ep\en)\KS}}$, respectively.

\begin{figure}[tbh]
    \centering
    \includegraphics[width=8.6cm]{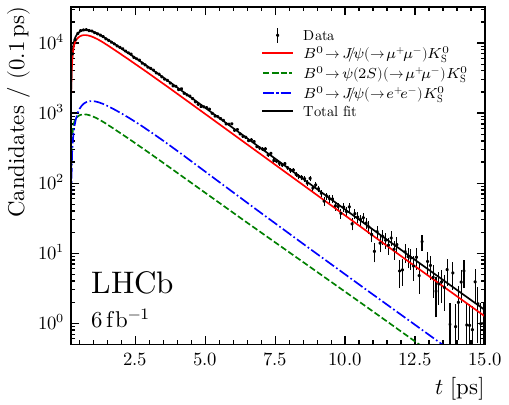}
    \caption{Decay-time distribution of the signal with an identified flavor at production, where the background is statistically subtracted by means of the sPlot method.
    The projections of the time-dependent fit for the individual contributions of the three decay modes and for the total are superimposed.}
    \label{fig:decay}
\end{figure}

\begin{figure}[tbh]
  \centering
  \includegraphics[width=8.6cm]{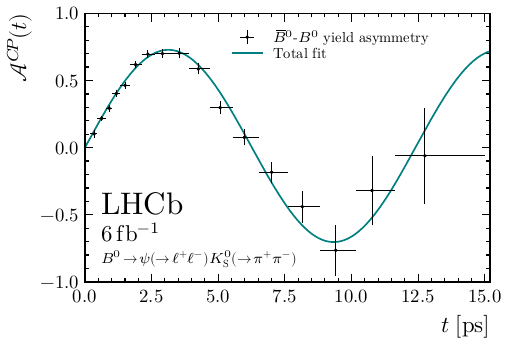}
  \caption{Time-dependent \CP asymmetry from the maximum-likelihood estimator of the binned asymmetry with the fit result overlaid.}
  \label{fig:yieldasymmetry}
\end{figure}

The \CP-violation parameters $S$ and $C$ are determined from a weighted maximum-likelihood fit to the time-dependent decay rates of \Bz and \Bzb tagged decays in the individual decay channels, after the results of individual years and reconstruction categories were found to be consistent.
In addition, a simultaneous fit of all channels is performed using the same model.
The time-dependent decay rate is expressed as
\begin{equation}
    \mathcal{P}(t, d, \eta)\propto e^{-\Gamma_\dquark t/\hbar}\bigg\{\big[1+d(1-2\omega^+(\eta))\big]P_\Bz(t)+\big[1+d(1-2\omega^-(\eta))\big]P_\Bzb(t)\bigg\}
    \label{eqn:cppdf}
\end{equation}
with
\begin{align}
    P_{\Bz,(\Bzb)}(t)&\propto(1\mp\alpha)(1\mp\Delta\epsilon_\text{tag})[1\mp S\sin(\dmd t)\pm C\cos(\dmd t)],
\end{align}
and where $\Gamma_\dquark$ is the \Bz decay width.
The decay-time resolution is accounted for in the fit by convolving Eq.~\eqref{eqn:cppdf} with a decay-time resolution model that is validated on simulation, corresponding to an effective resolution of about $60\fs$.
The effect of the resolution on the \CP-asymmetry amplitude is at the level of \SI{0.5}{\text{\textperthousand}} and thus small compared to the statistical sensitivity.
The resolution model consists of the sum of three Gaussian distributions centered at zero with widths defined as linear functions of the decay-time uncertainty.
A possible bias in the decay-time reconstruction due to \velo misalignment is considered and modeled by a second-degree polynomial of the decay-time uncertainty.
The parameters of the model are determined from a fit to the decay-time distribution of a data sample made of $\jpsi\pip\pim$ candidates compatible with originating from the PV.
The effect of the decay-time dependent signal-reconstruction efficiency is accounted for by multiplying the total PDF by a cubic spline model, whose shape is allowed to float in the fit.
The parameters \dmd and $\Gamma_\dquark$ are allowed to vary in the fit with Gaussian constraints to their known values~\cite{PDG2022}.
Similarly, the FT calibration parameters and the production asymmetry are constrained to the \decay{\Bz}{\jpsi\Kstarz} fit results using the full covariance matrix.
The effect of kaon regeneration and \CP violation in the neutral kaon system on the \CP-violation parameters of the \Bz system are estimated~\cite{Fetscher,Ko} and applied as a correction for each mode.
The correction assigned to the combined fit is $+0.0016$ for $S$ and $-0.0035$ for $C$.
Figure \ref{fig:decay} shows the decay-time distribution of the signal candidates with the fit result overlaid.
Figure \ref{fig:yieldasymmetry} shows the corresponding \CP asymmetry as a function of decay time, where the data points correspond to the maximum-likelihood estimator of the time-integrated \CP asymmetry in each decay-time bin, defined as ${\mathcal{A}^\CP_\text{int}=-(\sum_j^N\kappa_jd_jD_j)/(\sum_j^N\kappa_jD_j^2)}$, whereby $D_j=(1-\omega^+_j-\omega^-_j)$ is the tagging dilution, $d_j$ is the tagging decision, and $\kappa_j$ is the signal event weight obtained with the sPlot method.

Several sources of systematic uncertainties on the \CP-violating observables are investigated, including those associated with the choice of fit model and the uncertainty of the external inputs.
The corresponding effects are studied using pseudoexperiments in which ensembles of pseudodata are generated using parameters that differ from those used in the baseline fit.
The generated datasets are then fitted with the nominal model to test whether biases in the parameters of interest occur.
Each contribution is evaluated separately in each signal mode.
Sources of leading systematic uncertainty are listed in Table\ \ref{tab:systematics}.
A small bias in the result of the baseline fit is observed and assumed to be fully correlated among different signal modes.
The resulting systematic uncertainty on the combined result is obtained from the arithmetic mean of individual decay channels.
Other sources of systematic uncertainty are assumed to be uncorrelated.
The total systematic uncertainty for the combined fit is a weighted average of the individual uncertainties, taking into account the sensitivity of each mode to the \CP-violating parameters.
\begin{table}
    \centering
    \caption{Summary of systematic uncertainties on $S$ and $C$. Each contribution is a weighted average of the uncertainties of the individual fits, except for the fitter validation. }
    \label{tab:systematics}
    \begin{tabular}{ccc}
        \toprule
        Source & $\sigma(S)$ & $\sigma(C)$ \\ 
        \midrule
        Fitter validation                          & 0.0004 & 0.0006 \\
        Decay-time bias model                      & 0.0007 & 0.0013 \\
        FT $\Delta\epsilon_\text{tag}$ portability & 0.0014 & 0.0017 \\
        FT calibration portability                 & 0.0053 & 0.0001 \\
        $\DGd$ uncertainty                         & 0.0055 & 0.0017 \\
        \bottomrule
    \end{tabular}
\end{table}
The systematic uncertainty due to the uncertainty of \DGd is obtained by analyzing pseudoexperiments generated with input values shifted by $+1\sigma$ and $-1\sigma$ from its assumed central value of zero, where $\sigma$ is the sum of the current world average uncertainty and absolute central value.
The larger deviation is chosen as the systematic uncertainty.
Small differences in the FT calibration parameters and in the tagging efficiency asymmetry between the control channels and the signal modes are observed in simulation (FT calibration portability).
The contribution due to the portability of the FT calibration is evaluated by analyzing pseudoexperiments accounting for the different calibration values at generation level.
The systematic uncertainty due to the observed $\Delta\epsilon_\text{tag}$ differences in simulation is evaluated from the fit to data with modified input values.
Uncertainties on the parameters of the decay-time bias correction function are evaluated with pseudoexperiments, where the parameters are varied within their uncertainties.
The uncertainty of \dmd is included in the statistical uncertainty and its isolated contribution to the combined fit is $\sigma_\dmd(S)=0.0004$ and $\sigma_\dmd(C)=0.0023$.
The systematic uncertainty due to the choice of the sPlot method as background subtraction method is included in the fitter validation uncertainty.
Additional possible sources of systematic uncertainties are considered and found negligible.
These include the decay-time dependence of the FT efficiency and of the mistag; the choice of model for the decay-time efficiency, and the validation of the decay-time bias correction method.

Several cross-checks are performed to assess the consistency of the results by splitting the data by \KS reconstruction categories, years of data taking, SS and OS tagging amongst several others.
In addition, the fits are performed in bins of \PB-meson momentum, pseudorapidity, and other variables correlated to FT performance.
The results are also stable against variations of the mass-fit model and the choice of figure of merit used to determine the optimal BDT requirement.
Furthermore, the result is unaffected by the specific description of the decay-time efficiency.
The decay-time resolution and its modeling only play a minor role in this measurement due to the high decay-time resolution of the \lhcb detector and the comparatively slow \Bz oscillation.
The choice of decay-time resolution model therefore does not affect the result.
The influence of the correlation of invariant mass and decay time was found to be negligible.
Several alternative approaches to determine \CP-violating parameters are also performed and are consistent with the baseline results.
These include the determination of $S$ and $C$ parameters from the time-integrated asymmetry, a fit based on different assumptions on the normalizations of the \Bz and \Bzb amplitudes, and a statistical method to determine \CP-violation parameters using a model-independent approach to account for the decay-time efficiency.

\newpage
The \CP-violation parameters are measured to be
\begin{center}
    \renewcommand{\arraystretch}{1.2}
    \begin{tabular}{r@{$\;=\,$}S[table-format=-1.3]@{$\,\pm\,$}S[table-format=1.3]@{\,}l@{$\,\pm\,$}S[table-format=1.3]@{\,}l}
      $S_{\jpsi(\to\mup\mun)\KS}$    &  0.716  & 0.015 & (\text{stat}) & 0.007 & (\text{syst})\,, \\
      $C_{\jpsi(\to\mup\mun)\KS}$    &  0.010  & 0.014 & (\text{stat}) & 0.003 & (\text{syst})\,, \\
      $S_{\psitwos(\to\mup\mun)\KS}$ &  0.649  & 0.053 & (\text{stat}) & 0.018 & (\text{syst})\,, \\
      $C_{\psitwos(\to\mup\mun)\KS}$ & -0.087  & 0.048 & (\text{stat}) & 0.005 & (\text{syst})\,, \\
      $S_{\jpsi(\to\ep\en)\KS}$      &  0.754  & 0.037 & (\text{stat}) & 0.008 & (\text{syst})\,, \\
      $C_{\jpsi(\to\ep\en)\KS}$      &  0.042  & 0.034 & (\text{stat}) & 0.008 & (\text{syst})\,.
    \end{tabular}
\end{center}
The corresponding correlation coefficients between $S$ and $C$ are $0.446$, $0.503$, and $0.374$ for $\jpsi(\to\mup\mun)$, \psitwos(\to\mup\mun), and $\jpsi(\to\ep\en)$ final states, respectively.
A combined fit of the \decay{\Bz}{\jpsi(\mu\mu)\KS} and \decay{\Bz}{\jpsi(\to\electron\electron)\KS} modes results in
\begin{center}
    \renewcommand{\arraystretch}{1.2}
    \begin{tabular}{r@{$\,=\,$}S[table-format=1.3]@{$\,\pm\,$}S[table-format=1.3]@{\,}l@{$\,\pm\,$}S[table-format=1.3]@{\,}l}
      $S_{\jpsi\KS}$  & 0.722 & 0.014 & (\text{stat}) & 0.007 & (\text{syst})\,, \\
      $C_{\jpsi\KS}$  & 0.015 & 0.013 & (\text{stat}) & 0.003 & (\text{syst})\,,
    \end{tabular}
\end{center}
with a correlation coefficient of $0.437$.
A simultaneous fit of the three decay modes is performed and results in
\begin{center}
    \renewcommand{\arraystretch}{1.2}
    \begin{tabular}{r@{$\,=\,$}S[table-format=1.3]@{$\,\pm\,$}S[table-format=1.3]@{\,}l@{$\,\pm\,$}S[table-format=1.3]@{\,}l}
      $S_{\psi\KS}$  & 0.717 & 0.013 & (\text{stat}) & 0.008 & (\text{syst})\,, \\
      $C_{\psi\KS}$  & 0.008 & 0.012 & (\text{stat}) & 0.003 & (\text{syst})\,,
    \end{tabular}
\end{center}
with a correlation coefficient of $0.441$.
Finally, a combination of the \lhcb \runone and \runtwo results is performed.
It is assumed that sources 
of systematic uncertainties from external parameters \dmd, \DGd, and $\alpha$ are fully correlated between these measurements.
The combination of measurements yields
\begin{center}
    \renewcommand{\arraystretch}{1.2}
    \begin{tabular}{r@{$\,=\,$}S[table-format=1.3]@{$\,\pm\,$}S[table-format=1.3]@{\,}l}
      $S_{\psi\KS}^\runonetwo$  & 0.724 & 0.014 & (\text{stat+syst})\,, \\
      $C_{\psi\KS}^\runonetwo$  & 0.004 & 0.012 & (\text{stat+syst})\,,
    \end{tabular}
\end{center}
with a correlation coefficient of $0.40$, and for final states that contain a \jpsi meson the combination is
\begin{center}
    \renewcommand{\arraystretch}{1.2}
    \begin{tabular}{r@{$\,=\,$}S[table-format=1.3]@{$\,\pm\,$}S[table-format=1.3]@{\,}l}
      $S_{\jpsi\KS}^\runonetwo$  & 0.726 & 0.014 & (\text{stat+syst})\,, \\
      $C_{\jpsi\KS}^\runonetwo$  & 0.010 & 0.012 & (\text{stat+syst})\,,
    \end{tabular}
\end{center}
with a correlation coefficient of $0.41$.

In summary, a measurement of time-dependent \CP violation in ${\decay{\Bz}{\jpsi(\to\mup\mun)\KS}}$, ${\decay{\Bz}{\psitwos(\to\mup\mun)\KS}}$, and ${\decay{\Bz}{\jpsi(\to\ep\en)\KS}}$ decays using the full \lhcb \runtwo data corresponding to an integrated luminosity of 6\invfb from $\proton\proton$ collisions at a center-of-mass energy of 13\tev is reported.
These measurements are in agreement with recent predictions by the CKMfitter group~\cite{CKMfitter2015} and the UTfit group~\cite{UTfit-UT} and the current world averages as reported by HFLAV~\cite{HFLAV21}.
The result of the simultaneous fit to all channels is more precise than the current HFLAV world average.

\section*{Acknowledgements}
\noindent We express our gratitude to our colleagues in the CERN
accelerator departments for the excellent performance of the LHC. We
thank the technical and administrative staff at the LHCb
institutes.
We acknowledge support from CERN and from the national agencies:
CAPES, CNPq, FAPERJ and FINEP (Brazil); 
MOST and NSFC (China); 
CNRS/IN2P3 (France); 
BMBF, DFG and MPG (Germany); 
INFN (Italy); 
NWO (Netherlands); 
MNiSW and NCN (Poland); 
MCID/IFA (Romania); 
MICINN (Spain); 
SNSF and SER (Switzerland); 
NASU (Ukraine); 
STFC (United Kingdom); 
DOE NP and NSF (USA).
We acknowledge the computing resources that are provided by CERN, IN2P3
(France), KIT and DESY (Germany), INFN (Italy), SURF (Netherlands),
PIC (Spain), GridPP (United Kingdom), 
CSCS (Switzerland), IFIN-HH (Romania), CBPF (Brazil),
Polish WLCG  (Poland) and NERSC (USA).
We are indebted to the communities behind the multiple open-source
software packages on which we depend.
Individual groups or members have received support from
ARC and ARDC (Australia);
Minciencias (Colombia);
AvH Foundation (Germany);
EPLANET, Marie Sk\l{}odowska-Curie Actions, ERC and NextGenerationEU (European Union);
A*MIDEX, ANR, IPhU and Labex P2IO, and R\'{e}gion Auvergne-Rh\^{o}ne-Alpes (France);
Key Research Program of Frontier Sciences of CAS, CAS PIFI, CAS CCEPP, 
Fundamental Research Funds for the Central Universities, 
and Sci. \& Tech. Program of Guangzhou (China);
GVA, XuntaGal, GENCAT, Inditex, InTalent and Prog.~Atracci\'on Talento, CM (Spain);
SRC (Sweden);
the Leverhulme Trust, the Royal Society
 and UKRI (United Kingdom).

\addcontentsline{toc}{section}{References}

\ifx\mcitethebibliography\mciteundefinedmacro
\PackageError{LHCb.bst}{mciteplus.sty has not been loaded}
{This bibstyle requires the use of the mciteplus package.}\fi
\providecommand{\href}[2]{#2}
\begin{mcitethebibliography}{10}
\mciteSetBstSublistMode{n}
\mciteSetBstMaxWidthForm{subitem}{\alph{mcitesubitemcount})}
\mciteSetBstSublistLabelBeginEnd{\mcitemaxwidthsubitemform\space}
{\relax}{\relax}

\bibitem{HFLAV21}
Y.~Amhis {\em et~al.}, \ifthenelse{\boolean{articletitles}}{\emph{{Averages of
  $b$-hadron, $c$-hadron, and $\tau$-lepton properties as of 2021}},
  }{}\href{https://doi.org/10.1103/PhysRevD.107.052008}{Phys.\ Rev.\
  \textbf{D107} (2023) 052008},
  \href{http://arxiv.org/abs/2206.07501}{{\normalfont\ttfamily
  arXiv:2206.07501}}, {updated results and plots available at
  \href{https://hflav.web.cern.ch}{{\texttt{https://hflav.web.cern.ch}}}}\relax
\mciteBstWouldAddEndPuncttrue
\mciteSetBstMidEndSepPunct{\mcitedefaultmidpunct}
{\mcitedefaultendpunct}{\mcitedefaultseppunct}\relax
\EndOfBibitem
\bibitem{PhysRevLett.87.091801}
BABAR collaboration, B.~Aubert {\em et~al.},
  \ifthenelse{\boolean{articletitles}}{\emph{Observation of {\CP} violation in
  the {\Bz} meson system},
  }{}\href{https://doi.org/10.1103/PhysRevLett.87.091801}{Phys.\ Rev.\ Lett.\
  \textbf{87} (2001) 091801}\relax
\mciteBstWouldAddEndPuncttrue
\mciteSetBstMidEndSepPunct{\mcitedefaultmidpunct}
{\mcitedefaultendpunct}{\mcitedefaultseppunct}\relax
\EndOfBibitem
\bibitem{PhysRevLett.87.091802}
Belle collaboration, K.~Abe {\em et~al.},
  \ifthenelse{\boolean{articletitles}}{\emph{Observation of large {\CP}
  violation in the neutral {\PB} meson system},
  }{}\href{https://doi.org/10.1103/PhysRevLett.87.091802}{Phys.\ Rev.\ Lett.\
  \textbf{87} (2001) 091802}\relax
\mciteBstWouldAddEndPuncttrue
\mciteSetBstMidEndSepPunct{\mcitedefaultmidpunct}
{\mcitedefaultendpunct}{\mcitedefaultseppunct}\relax
\EndOfBibitem
\bibitem{BaBar:2009byl}
BABAR collaboration, B.~Aubert {\em et~al.},
  \ifthenelse{\boolean{articletitles}}{\emph{Measurement of time-dependent
  {\CP} asymmetry in {$\decay{\Bz}{\cquark\cquarkbar \kaon^{(*)0}}$} decays},
  }{}\href{https://doi.org/10.1103/PhysRevD.79.072009}{Phys.\ Rev.\
  \textbf{D79} (2009) 072009},
  \href{http://arxiv.org/abs/0902.1708}{{\normalfont\ttfamily
  arXiv:0902.1708}}\relax
\mciteBstWouldAddEndPuncttrue
\mciteSetBstMidEndSepPunct{\mcitedefaultmidpunct}
{\mcitedefaultendpunct}{\mcitedefaultseppunct}\relax
\EndOfBibitem
\bibitem{Belle:2012paq}
Belle collaboration, I.~Adachi {\em et~al.},
  \ifthenelse{\boolean{articletitles}}{\emph{Precise measurement of the {\CP}
  violation parameter {$\sin 2\phi_1$} in {\decay{\Bz}{(\cquark\cquarkbar)\Kz}}
  decays}, }{}\href{https://doi.org/10.1103/PhysRevLett.108.171802}{Phys.\
  Rev.\ Lett.\  \textbf{108} (2012) 171802},
  \href{http://arxiv.org/abs/1201.4643}{{\normalfont\ttfamily
  arXiv:1201.4643}}\relax
\mciteBstWouldAddEndPuncttrue
\mciteSetBstMidEndSepPunct{\mcitedefaultmidpunct}
{\mcitedefaultendpunct}{\mcitedefaultseppunct}\relax
\EndOfBibitem
\bibitem{LHCb-PAPER-2015-004}
LHCb collaboration, R.~Aaij {\em et~al.},
  \ifthenelse{\boolean{articletitles}}{\emph{{Measurement of \CP violation in
  \mbox{\decay{\Bz}{\jpsi\KS}} decays}},
  }{}\href{https://doi.org/10.1103/PhysRevLett.115.031601}{Phys.\ Rev.\ Lett.\
  \textbf{115} (2015) 031601},
  \href{http://arxiv.org/abs/1503.07089}{{\normalfont\ttfamily
  arXiv:1503.07089}}\relax
\mciteBstWouldAddEndPuncttrue
\mciteSetBstMidEndSepPunct{\mcitedefaultmidpunct}
{\mcitedefaultendpunct}{\mcitedefaultseppunct}\relax
\EndOfBibitem
\bibitem{LHCb-PAPER-2017-029}
LHCb collaboration, R.~Aaij {\em et~al.},
  \ifthenelse{\boolean{articletitles}}{\emph{{Measurement of \CP violation in
  \mbox{\decay{\Bz}{\jpsi \KS}} and \mbox{\decay{\Bz}{\psitwos \KS}} decays}},
  }{}\href{https://doi.org/10.1007/JHEP11(2017)170}{JHEP \textbf{11} (2017)
  170}, \href{http://arxiv.org/abs/1709.03944}{{\normalfont\ttfamily
  arXiv:1709.03944}}\relax
\mciteBstWouldAddEndPuncttrue
\mciteSetBstMidEndSepPunct{\mcitedefaultmidpunct}
{\mcitedefaultendpunct}{\mcitedefaultseppunct}\relax
\EndOfBibitem
\bibitem{Belle-II:2023nmj}
Belle II collaboration, I.~Adachi {\em et~al.},
  \ifthenelse{\boolean{articletitles}}{\emph{Measurement of
  decay-time-dependent {\CP} violation in \decay{\Bz}{\jpsi\KS} decays using
  2019-2021 {Belle II} data},
  }{}\href{http://arxiv.org/abs/2302.12898}{{\normalfont\ttfamily
  arXiv:2302.12898}}\relax
\mciteBstWouldAddEndPuncttrue
\mciteSetBstMidEndSepPunct{\mcitedefaultmidpunct}
{\mcitedefaultendpunct}{\mcitedefaultseppunct}\relax
\EndOfBibitem
\bibitem{CPreviewChap}
T.~Gershon and Y.~Nir, \ifthenelse{\boolean{articletitles}}{\emph{{\CP}
  violation in the quark sector}, }{} 2022.
\newblock review in Ref.~\cite{PDG2022}\relax
\mciteBstWouldAddEndPuncttrue
\mciteSetBstMidEndSepPunct{\mcitedefaultmidpunct}
{\mcitedefaultendpunct}{\mcitedefaultseppunct}\relax
\EndOfBibitem
\bibitem{PDG2022}
Particle Data Group, R.~L. Workman {\em et~al.},
  \ifthenelse{\boolean{articletitles}}{\emph{{\href{http://pdg.lbl.gov/}{Review
  of particle physics}}}, }{}\href{https://doi.org/10.1093/ptep/ptac097}{Prog.\
  Theor.\ Exp.\ Phys.\  \textbf{2022} (2022) 083C01}\relax
\mciteBstWouldAddEndPuncttrue
\mciteSetBstMidEndSepPunct{\mcitedefaultmidpunct}
{\mcitedefaultendpunct}{\mcitedefaultseppunct}\relax
\EndOfBibitem
\bibitem{DeBruyn:2014oga}
K.~De~Bruyn and R.~Fleischer, \ifthenelse{\boolean{articletitles}}{\emph{A
  roadmap to control penguin effects in {$\decay{B^0_\dquark}{\jpsi\KS}$} and
  {$\decay{B^0_\squark}{ \jpsi\phi}$}},
  }{}\href{https://doi.org/10.1007/JHEP03(2015)145}{JHEP \textbf{03} (2015)
  145}, \href{http://arxiv.org/abs/1412.6834}{{\normalfont\ttfamily
  arXiv:1412.6834}}\relax
\mciteBstWouldAddEndPuncttrue
\mciteSetBstMidEndSepPunct{\mcitedefaultmidpunct}
{\mcitedefaultendpunct}{\mcitedefaultseppunct}\relax
\EndOfBibitem
\bibitem{Barel:2020jvf}
M.~Z. Barel, K.~De~Bruyn, R.~Fleischer, and E.~Malami,
  \ifthenelse{\boolean{articletitles}}{\emph{In pursuit of new physics with
  {$\decay{\PB_{\dquark}^0}{\jpsi\Kz}$} and {$\decay{\Bs}{\jpsi\phi}$} decays
  at the high-precision frontier},
  }{}\href{https://doi.org/10.1088/1361-6471/abf2a2}{J.\ Phys.\  \textbf{G48}
  (2021) 065002}, \href{http://arxiv.org/abs/2010.14423}{{\normalfont\ttfamily
  arXiv:2010.14423}}\relax
\mciteBstWouldAddEndPuncttrue
\mciteSetBstMidEndSepPunct{\mcitedefaultmidpunct}
{\mcitedefaultendpunct}{\mcitedefaultseppunct}\relax
\EndOfBibitem
\bibitem{Ciuchini:2005mg}
M.~Ciuchini, M.~Pierini, and L.~Silvestrini,
  \ifthenelse{\boolean{articletitles}}{\emph{Effect of penguin operators in the
  \decay{\Bz}{\jpsi\Kz} {\CP} asymmetry},
  }{}\href{https://doi.org/10.1103/PhysRevLett.95.221804}{Phys.\ Rev.\ Lett.\
  \textbf{95} (2005) 221804}\relax
\mciteBstWouldAddEndPuncttrue
\mciteSetBstMidEndSepPunct{\mcitedefaultmidpunct}
{\mcitedefaultendpunct}{\mcitedefaultseppunct}\relax
\EndOfBibitem
\bibitem{Faller:2008zc}
S.~Faller, M.~Jung, R.~Fleischer, and T.~Mannel,
  \ifthenelse{\boolean{articletitles}}{\emph{The golden modes
  {$\decay{\Bz}{\jpsi\kaon_{\mathrm{S,L}}}$} in the era of precision flavour
  physics}, }{}\href{https://doi.org/10.1103/PhysRevD.79.014030}{Phys.\ Rev.\
  \textbf{D79} (2009) 014030},
  \href{http://arxiv.org/abs/0809.0842}{{\normalfont\ttfamily
  arXiv:0809.0842}}\relax
\mciteBstWouldAddEndPuncttrue
\mciteSetBstMidEndSepPunct{\mcitedefaultmidpunct}
{\mcitedefaultendpunct}{\mcitedefaultseppunct}\relax
\EndOfBibitem
\bibitem{Jung:2012mp}
M.~Jung, \ifthenelse{\boolean{articletitles}}{\emph{Determining weak phases
  from {\decay{\PB}{\jpsi\PP}} decays},
  }{}\href{https://doi.org/10.1103/PhysRevD.86.053008}{Phys.\ Rev.\
  \textbf{D86} (2012) 053008},
  \href{http://arxiv.org/abs/1206.2050}{{\normalfont\ttfamily
  arXiv:1206.2050}}\relax
\mciteBstWouldAddEndPuncttrue
\mciteSetBstMidEndSepPunct{\mcitedefaultmidpunct}
{\mcitedefaultendpunct}{\mcitedefaultseppunct}\relax
\EndOfBibitem
\bibitem{Frings:2015eva}
P.~Frings, U.~Nierste, and M.~Wiebusch,
  \ifthenelse{\boolean{articletitles}}{\emph{Penguin contributions to {\CP}
  phases in {$\PB_{\dquark,\squark}$} decays to charmonium},
  }{}\href{https://doi.org/10.1103/PhysRevLett.115.061802}{Phys.\ Rev.\ Lett.\
  \textbf{115} (2015) 061802},
  \href{http://arxiv.org/abs/1503.00859}{{\normalfont\ttfamily
  arXiv:1503.00859}}\relax
\mciteBstWouldAddEndPuncttrue
\mciteSetBstMidEndSepPunct{\mcitedefaultmidpunct}
{\mcitedefaultendpunct}{\mcitedefaultseppunct}\relax
\EndOfBibitem
\bibitem{LHCb-DP-2008-001}
LHCb collaboration, A.~A. Alves~Jr.\ {\em et~al.},
  \ifthenelse{\boolean{articletitles}}{\emph{{The \lhcb detector at the LHC}},
  }{}\href{https://doi.org/10.1088/1748-0221/3/08/S08005}{JINST \textbf{3}
  (2008) S08005}\relax
\mciteBstWouldAddEndPuncttrue
\mciteSetBstMidEndSepPunct{\mcitedefaultmidpunct}
{\mcitedefaultendpunct}{\mcitedefaultseppunct}\relax
\EndOfBibitem
\bibitem{LHCb-DP-2014-002}
LHCb collaboration, R.~Aaij {\em et~al.},
  \ifthenelse{\boolean{articletitles}}{\emph{{LHCb detector performance}},
  }{}\href{https://doi.org/10.1142/S0217751X15300227}{Int.\ J.\ Mod.\ Phys.\
  \textbf{A30} (2015) 1530022},
  \href{http://arxiv.org/abs/1412.6352}{{\normalfont\ttfamily
  arXiv:1412.6352}}\relax
\mciteBstWouldAddEndPuncttrue
\mciteSetBstMidEndSepPunct{\mcitedefaultmidpunct}
{\mcitedefaultendpunct}{\mcitedefaultseppunct}\relax
\EndOfBibitem
\bibitem{LHCb-DP-2012-004}
R.~Aaij {\em et~al.}, \ifthenelse{\boolean{articletitles}}{\emph{{The \lhcb
  trigger and its performance in 2011}},
  }{}\href{https://doi.org/10.1088/1748-0221/8/04/P04022}{JINST \textbf{8}
  (2013) P04022}, \href{http://arxiv.org/abs/1211.3055}{{\normalfont\ttfamily
  arXiv:1211.3055}}\relax
\mciteBstWouldAddEndPuncttrue
\mciteSetBstMidEndSepPunct{\mcitedefaultmidpunct}
{\mcitedefaultendpunct}{\mcitedefaultseppunct}\relax
\EndOfBibitem
\bibitem{XGBoost}
T.~Chen and C.~Guestrin, \ifthenelse{\boolean{articletitles}}{\emph{{XGBoost
  (version 1.0.2)}: A scalable tree boosting system}, }{} in {\em Proceedings
  of the 22nd ACM SIGKDD International Conference on Knowledge Discovery and
  Data Mining} KDD '16 (2016) 785--794, \href{https://arxiv.org/abs/1603.02754}{{\normalfont\ttfamily arXiv:1603.02754}}\relax
\mciteBstWouldAddEndPuncttrue
\mciteSetBstMidEndSepPunct{\mcitedefaultmidpunct}
{\mcitedefaultendpunct}{\mcitedefaultseppunct}\relax
\EndOfBibitem
\bibitem{Sjostrand:2007gs}
T.~Sj\"{o}strand, S.~Mrenna, and P.~Skands,
  \ifthenelse{\boolean{articletitles}}{\emph{{A brief introduction to PYTHIA
  8.1}}, }{}\href{https://doi.org/10.1016/j.cpc.2008.01.036}{Comput.\ Phys.\
  Commun.\  \textbf{178} (2008) 852},
  \href{http://arxiv.org/abs/0710.3820}{{\normalfont\ttfamily
  arXiv:0710.3820}}\relax
\mciteBstWouldAddEndPuncttrue
\mciteSetBstMidEndSepPunct{\mcitedefaultmidpunct}
{\mcitedefaultendpunct}{\mcitedefaultseppunct}\relax
\EndOfBibitem
\bibitem{Sjostrand:2006za}
T.~Sj\"{o}strand, S.~Mrenna, and P.~Skands,
  \ifthenelse{\boolean{articletitles}}{\emph{{PYTHIA 6.4 physics and manual}},
  }{}\href{https://doi.org/10.1088/1126-6708/2006/05/026}{JHEP \textbf{05}
  (2006) 026}, \href{http://arxiv.org/abs/hep-ph/0603175}{{\normalfont\ttfamily
  arXiv:hep-ph/0603175}}\relax
\mciteBstWouldAddEndPuncttrue
\mciteSetBstMidEndSepPunct{\mcitedefaultmidpunct}
{\mcitedefaultendpunct}{\mcitedefaultseppunct}\relax
\EndOfBibitem
\bibitem{LHCb-PROC-2010-056}
I.~Belyaev {\em et~al.}, \ifthenelse{\boolean{articletitles}}{\emph{{Handling
  of the generation of primary events in Gauss, the LHCb simulation
  framework}}, }{}\href{https://doi.org/10.1088/1742-6596/331/3/032047}{J.\
  Phys.\ Conf.\ Ser.\  \textbf{331} (2011) 032047}\relax
\mciteBstWouldAddEndPuncttrue
\mciteSetBstMidEndSepPunct{\mcitedefaultmidpunct}
{\mcitedefaultendpunct}{\mcitedefaultseppunct}\relax
\EndOfBibitem
\bibitem{Allison:2006ve}
Geant4 collaboration, J.~Allison {\em et~al.},
  \ifthenelse{\boolean{articletitles}}{\emph{{Geant4 developments and
  applications}}, }{}\href{https://doi.org/10.1109/TNS.2006.869826}{IEEE
  Trans.\ Nucl.\ Sci.\  \textbf{53} (2006) 270}\relax
\mciteBstWouldAddEndPuncttrue
\mciteSetBstMidEndSepPunct{\mcitedefaultmidpunct}
{\mcitedefaultendpunct}{\mcitedefaultseppunct}\relax
\EndOfBibitem
\bibitem{Agostinelli:2002hh}
Geant4 collaboration, S.~Agostinelli {\em et~al.},
  \ifthenelse{\boolean{articletitles}}{\emph{{Geant4: A simulation toolkit}},
  }{}\href{https://doi.org/10.1016/S0168-9002(03)01368-8}{Nucl.\ Instrum.\
  Meth.\  \textbf{A506} (2003) 250}\relax
\mciteBstWouldAddEndPuncttrue
\mciteSetBstMidEndSepPunct{\mcitedefaultmidpunct}
{\mcitedefaultendpunct}{\mcitedefaultseppunct}\relax
\EndOfBibitem
\bibitem{Lange:2001uf}
D.~J. Lange, \ifthenelse{\boolean{articletitles}}{\emph{{The EvtGen particle
  decay simulation package}},
  }{}\href{https://doi.org/10.1016/S0168-9002(01)00089-4}{Nucl.\ Instrum.\
  Meth.\  \textbf{A462} (2001) 152}\relax
\mciteBstWouldAddEndPuncttrue
\mciteSetBstMidEndSepPunct{\mcitedefaultmidpunct}
{\mcitedefaultendpunct}{\mcitedefaultseppunct}\relax
\EndOfBibitem
\bibitem{davidson2015photos}
N.~Davidson, T.~Przedzinski, and Z.~Was,
  \ifthenelse{\boolean{articletitles}}{\emph{{PHOTOS interface in C++:
  Technical and physics documentation}},
  }{}\href{https://doi.org/https://doi.org/10.1016/j.cpc.2015.09.013}{Comp.\
  Phys.\ Comm.\  \textbf{199} (2016) 86},
  \href{http://arxiv.org/abs/1011.0937}{{\normalfont\ttfamily
  arXiv:1011.0937}}\relax
\mciteBstWouldAddEndPuncttrue
\mciteSetBstMidEndSepPunct{\mcitedefaultmidpunct}
{\mcitedefaultendpunct}{\mcitedefaultseppunct}\relax
\EndOfBibitem
\bibitem{Blum1999BeatingTH}
A.~Blum, A.~T. Kalai, and J.~Langford,
  \ifthenelse{\boolean{articletitles}}{\emph{Beating the hold-out: bounds for
  k-fold and progressive cross-validation}, }{} in {\em Proceedings of the twelfth annual conference on computational learning theory}, \href{https://dl.acm.org/doi/10.1145/307400.307439}{Association for Computing Machinery, New York, 1999}\relax
\mciteBstWouldAddEndPuncttrue
\mciteSetBstMidEndSepPunct{\mcitedefaultmidpunct}
{\mcitedefaultendpunct}{\mcitedefaultseppunct}\relax
\EndOfBibitem
\bibitem{LHCb-PAPER-2011-027}
LHCb collaboration, R.~Aaij {\em et~al.},
  \ifthenelse{\boolean{articletitles}}{\emph{{Opposite-side flavour tagging of
  \B mesons at the LHCb experiment}},
  }{}\href{https://doi.org/10.1140/epjc/s10052-012-2022-1}{Eur.\ Phys.\ J.\
  \textbf{C72} (2012) 2022},
  \href{http://arxiv.org/abs/1202.4979}{{\normalfont\ttfamily
  arXiv:1202.4979}}\relax
\mciteBstWouldAddEndPuncttrue
\mciteSetBstMidEndSepPunct{\mcitedefaultmidpunct}
{\mcitedefaultendpunct}{\mcitedefaultseppunct}\relax
\EndOfBibitem
\bibitem{Fazzini:2018dyq}
D.~Fazzini, \ifthenelse{\boolean{articletitles}}{\emph{{Flavour Tagging in the
  LHCb experiment}}, }{} in {\em {Proceedings, 6th Large Hadron Collider
  Physics Conference (LHCP 2018): Bologna, Italy, 2018}},
  \href{https://doi.org/10.22323/1.321.0230}{Proceedings of Science (PoS), Bologna, Italy, 2018 \textbf{LHCP2018} 230}\relax
\mciteBstWouldAddEndPuncttrue
\mciteSetBstMidEndSepPunct{\mcitedefaultmidpunct}
{\mcitedefaultendpunct}{\mcitedefaultseppunct}\relax
\EndOfBibitem
\bibitem{ftcalib}
Q.~Führing and V.~Jevti\'c,
  \ifthenelse{\boolean{articletitles}}{\emph{{lhcb-ftcalib (version 1.4.0)}:
  {\lhcb Flavour Tagging calibration software}}, }{} 2022.
\newblock \url{https://gitlab.cern.ch/lhcb-ft/lhcb_ftcalib}\relax
\mciteBstWouldAddEndPuncttrue
\mciteSetBstMidEndSepPunct{\mcitedefaultmidpunct}
{\mcitedefaultendpunct}{\mcitedefaultseppunct}\relax
\EndOfBibitem
\bibitem{Pivk:2004ty}
M.~Pivk and F.~R. Le~Diberder,
  \ifthenelse{\boolean{articletitles}}{\emph{{sPlot: A statistical tool to
  unfold data distributions}},
  }{}\href{https://doi.org/10.1016/j.nima.2005.08.106}{Nucl.\ Instrum.\ Meth.\
  \textbf{A555} (2005) 356},
  \href{http://arxiv.org/abs/physics/0402083}{{\normalfont\ttfamily
  arXiv:physics/0402083}}\relax
\mciteBstWouldAddEndPuncttrue
\mciteSetBstMidEndSepPunct{\mcitedefaultmidpunct}
{\mcitedefaultendpunct}{\mcitedefaultseppunct}\relax
\EndOfBibitem
\bibitem{Xie:2009rka}
Y.~Xie, \ifthenelse{\boolean{articletitles}}{\emph{{sFit: a method for
  background subtraction in maximum likelihood fit}},
  }{}\href{http://arxiv.org/abs/0905.0724}{{\normalfont\ttfamily
  arXiv:0905.0724}}\relax
\mciteBstWouldAddEndPuncttrue
\mciteSetBstMidEndSepPunct{\mcitedefaultmidpunct}
{\mcitedefaultendpunct}{\mcitedefaultseppunct}\relax
\EndOfBibitem
\bibitem{Santos:2013gra}
D.~Mart{\'\i}nez~Santos and F.~Dupertuis,
  \ifthenelse{\boolean{articletitles}}{\emph{{Mass distributions marginalized
  over per-event errors}},
  }{}\href{https://doi.org/10.1016/j.nima.2014.06.081}{Nucl.\ Instrum.\ Meth.\
  \textbf{A764} (2014) 150},
  \href{http://arxiv.org/abs/1312.5000}{{\normalfont\ttfamily
  arXiv:1312.5000}}\relax
\mciteBstWouldAddEndPuncttrue
\mciteSetBstMidEndSepPunct{\mcitedefaultmidpunct}
{\mcitedefaultendpunct}{\mcitedefaultseppunct}\relax
\EndOfBibitem
\bibitem{Fetscher}
W.~Fetscher {\em et~al.},
  \ifthenelse{\boolean{articletitles}}{\emph{Regeneration of arbitrary coherent
  neutral kaon states: A new method for measuring the {\Kz}-{\Kzb} forward
  scattering amplitude}, }{}\href{https://doi.org/10.1007/s002880050277}{Z.\
  Phys.\  \textbf{C72} (1996) 543}\relax
\mciteBstWouldAddEndPuncttrue
\mciteSetBstMidEndSepPunct{\mcitedefaultmidpunct}
{\mcitedefaultendpunct}{\mcitedefaultseppunct}\relax
\EndOfBibitem
\bibitem{Ko}
B.~R. Ko, E.~Won, B.~Golob, and P.~Pakhlov,
  \ifthenelse{\boolean{articletitles}}{\emph{Effect of nuclear interactions of
  neutral kaons on {\CP} asymmetry measurements},
  }{}\href{https://doi.org/10.1103/PhysRevD.84.111501}{Phys.\ Rev.\
  \textbf{D84} (2011) 111501},
  \href{http://arxiv.org/abs/1006.1938}{{\normalfont\ttfamily
  arXiv:1006.1938}}\relax
\mciteBstWouldAddEndPuncttrue
\mciteSetBstMidEndSepPunct{\mcitedefaultmidpunct}
{\mcitedefaultendpunct}{\mcitedefaultseppunct}\relax
\EndOfBibitem
\bibitem{CKMfitter2015}
CKMfitter group, J.~Charles {\em et~al.},
  \ifthenelse{\boolean{articletitles}}{\emph{{Current status of the standard
  model CKM fit and constraints on \hbox{$\Delta F=2$} new physics}},
  }{}\href{https://doi.org/10.1103/PhysRevD.91.073007}{Phys.\ Rev.\
  \textbf{D91} (2015) 073007},
  \href{http://arxiv.org/abs/1501.05013}{{\normalfont\ttfamily
  arXiv:1501.05013}}, {updated results and plots available at
  \href{http://ckmfitter.in2p3.fr/}{{\texttt{http://ckmfitter.in2p3.fr/}}}}\relax
\mciteBstWouldAddEndPuncttrue
\mciteSetBstMidEndSepPunct{\mcitedefaultmidpunct}
{\mcitedefaultendpunct}{\mcitedefaultseppunct}\relax
\EndOfBibitem
\bibitem{UTfit-UT}
UTfit collaboration, M.~Bona {\em et~al.},
  \ifthenelse{\boolean{articletitles}}{\emph{{The unitarity triangle fit in the
  standard model and hadronic parameters from lattice QCD: A reappraisal after
  the measurements of $\Delta m_{s}$ and $BR(B\to\tau\nu_{\tau})$}},
  }{}\href{https://doi.org/10.1088/1126-6708/2006/10/081}{JHEP \textbf{10}
  (2006) 081}, \href{http://arxiv.org/abs/hep-ph/0606167}{{\normalfont\ttfamily
  arXiv:hep-ph/0606167}}, {updated results and plots available at
  \href{http://www.utfit.org/}{{\texttt{http://www.utfit.org/}}}}\relax
\mciteBstWouldAddEndPuncttrue
\mciteSetBstMidEndSepPunct{\mcitedefaultmidpunct}
{\mcitedefaultendpunct}{\mcitedefaultseppunct}\relax
\EndOfBibitem
\end{mcitethebibliography}

\ifx\mcitethebibliography\mciteundefinedmacro
\PackageError{LHCb.bst}{mciteplus.sty has not been loaded}
{This bibstyle requires the use of the mciteplus package.}\fi
\providecommand{\href}[2]{#2}

\newpage

\centerline
{\large\bf LHCb collaboration}
\begin
{flushleft}
\small
R.~Aaij$^{33}$\lhcborcid{0000-0003-0533-1952},
A.S.W.~Abdelmotteleb$^{52}$\lhcborcid{0000-0001-7905-0542},
C.~Abellan~Beteta$^{46}$,
F.~Abudin{\'e}n$^{52}$\lhcborcid{0000-0002-6737-3528},
T.~Ackernley$^{56}$\lhcborcid{0000-0002-5951-3498},
B.~Adeva$^{42}$\lhcborcid{0000-0001-9756-3712},
M.~Adinolfi$^{50}$\lhcborcid{0000-0002-1326-1264},
P.~Adlarson$^{78}$\lhcborcid{0000-0001-6280-3851},
H.~Afsharnia$^{10}$,
C.~Agapopoulou$^{44}$\lhcborcid{0000-0002-2368-0147}
% ...

{\footnotesize \it

% ...
$^{10}$Universit{\'e} Clermont Auvergne, CNRS/IN2P3, LPC, Clermont-Ferrand, France\\
% ...
$^{33}$Nikhef National Institute for Subatomic Physics, Amsterdam, Netherlands\\
% ...
$^{42}$Instituto Galego de F{\'\i}sica de Altas Enerx{\'\i}as (IGFAE), Universidade de Santiago de Compostela, Santiago de Compostela, Spain\\
% ...
$^{44}$European Organization for Nuclear Research (CERN), Geneva, Switzerland\\
% ...
$^{46}$Physik-Institut, Universit{\"a}t Z{\"u}rich, Z{\"u}rich, Switzerland\\
% ...
$^{50}$H.H. Wills Physics Laboratory, University of Bristol, Bristol, United Kingdom\\
% ...
$^{52}$Department of Physics, University of Warwick, Coventry, United Kingdom\\
% ...
$^{55}$School of Physics and Astronomy, University of Glasgow, Glasgow, United Kingdom\\
$^{56}$Oliver Lodge Laboratory, University of Liverpool, Liverpool, United Kingdom\\
% ...
$^{78}$Department of Physics and Astronomy, Uppsala University, Uppsala, Sweden, associated to $^{55}$\\
% ...
}
\end{flushleft}

\end{document}